\documentclass[12pt]{article}   
 \usepackage{times,fancyheadings}
        
\begin{document}
\thispagestyle{empty} 

% \lhead[\fancyplain{}{\sl Price waves}]{\fancyplain{}{\sl The    
% Newsletter  of Econophysics}}
% \rhead[\fancyplain{}{\sl The Newsletter of Econophysics}]{\fancyplain{}{\sl   
% April 1997 Number 1}}

 \lhead[\fancyplain{}{\sl }]{\fancyplain{}{\sl }}
 \rhead[\fancyplain{}{\sl }]{\fancyplain{}{\sl }}

%%%%% Definitions

\newcommand{\nc}{\newcommand}
\newcommand \be{\begin{equation}}
\newcommand \bea{\begin{eqnarray} \nonumber }
\newcommand \ee{\end{equation}}
\newcommand \eea{\end{eqnarray}}
\

\nc{\qI}[1]{\section{ {#1} }}
\nc{\qA}[1]{\subsection{ {#1} }}
\nc{\qun}[1]{\subsubsection{ {#1} }}
\nc{\qa}[1]{\paragraph{ {#1} }}

\nc{\qfoot}[1]{\footnote{ {#1} }}
\def\qL{\hfill \break}
\def\qpar{\vskip 2mm plus 0.2mm minus 0.2mm}

\def\qparr{ \vskip 1.0mm plus 0.2mm minus 0.2mm \hangindent=10mm
\hangafter=1}

                % Decale UN paragraphe
                % Attention! La double accolade est vitale, sinon tout le
                % est decale (cf TEX p.199)
                % On peut aller a la ligne avec \qL=\hfill \break
                % Par contre ne supporte pas les lignes blanches
\def\qdec#1{\par {\leftskip=2cm {#1} \par}}

   %% Defs specifiques
\def\qdpt{\partial_t}
\def\qdpx{\partial_x}
\def\qddpt{\partial^{2}_{t^2}}
\def\qddpx{\partial^{2}_{x^2}}
\def\qn#1{\eqno \hbox{(#1)}}
\def\qds{\displaystyle}
\def\qal{\sqrt{1+\alpha ^2}}
\def\qw{\widetilde}

%%%%% End of definitions

\null
\vskip 1cm

\centerline{\bf \LARGE ``Thermometers'' of Speculative Frenzy}

\vskip 1cm
\centerline{\bf  B.M. Roehner$^1$ and D. Sornette$^2$}
\centerline{ $^1$ L.P.T.H.E., University Paris 7, 2 Place Jussieu}
\centerline{ 75251 Paris Cedex 05, France, roehner@lpthe.jussieu.fr}
\centerline{ $^2$ Institute of Geophysics and Planetary Physics 
and Department of Earth and Space Science}
\centerline{ University of California, Los Angeles, California 90095}
\centerline{ and Laboratoire de Physique de la Mati\`ere Condens\'ee}
\centerline{ CNRS UMR6622  and Universit\'e des Sciences}
\centerline{ B.P. 70, Parc Valrose, 06108 Nice Cedex 2, France}
\centerline{ sornette@moho.ess.ucla.edu}
\vskip 2cm

{\bf Abstract}~~~Establishing unambiguously the existence of speculative bubbles
is an on-going controversy complicated by the need of defining a model of fundamental prices.
Here, we present a novel empirical method which bypasses all the difficulties
of the previous approaches by monitoring external indicators of
an anomalously growing interest in the public at times of bubbles. 
From the definition of a bubble
as a self-fulfilling reinforcing price change, we identify
indicators of a possible self-reinforcing imitation
between agents in the market. We show that during the
build-up phase of a bubble, there is a growing interest in the public for the 
commodity in question, whether it consists in stocks, diamonds or coins. 
That interest can be estimated through different indicators: increase in
the number of books published on the topic, increase in the subscriptions
to specialized journals. Moreover, the well-known empirical rule according to
which the volume of sales is growing during a bull market finds a natural
interpretation in this framework: sales increases in fact reveal and
pinpoint the progress of the bubble's diffusion throughout society. 
We also present a simple model
of rational expectation which maps exactly onto the Ising model on a random graph.
The indicators are then interpreted as ``thermometers'', measuring the balance
between idiosyncratic information (noise temperature) and imitation (coupling) strength.
In this context, bubbles are interpreted as low or critical temperature phases,
where the imitation strength carries market prices up essentially independently of
fundamentals. Contrary to the naive conception of a bubble and a crash as times of
disorder, on the contrary, we show that bubbles and crashes are times where the concensus
is too strong!

\vskip 1cm
  \centerline{first version 14 Oct. 1999}
    \centerline{this version: 18 January 2000 }

\vskip 1cm
{\bf PACS.}\ 64.60 Equilibrium properties near critical points - 
 87.23Ge Dynamics of social systems 

\vfill \eject

\qI{The hidden collective factor of speculative bubbles}

In economic models, the price of a commodity is 
determined by its supply and demand functions. In the presence of
unexpected events, ``explanations'' are forged by mechanisms that appear 
to be similar in economics and in history.

\qA{``Explanations'' of unexpected events}

When the king of Sweden Charles XII was defeated by the Russians at the
battle of Poltava (1709), Europe was stunned for until then the Swedish
army was considered invincible. Subsequently, the defeat was ``explained''
in various ways. For instance, several historians 
(e.g. Andersson 1973, Scott 1977) attributed the defeat
to the fact that, having been wounded in the leg two days 
before, Charles XII could not lead his army. 
Even if this argument appears reasonable,
one can doubt its soundness when one notes the huge numerical superiority
of the Russians, a point which surprisingly enough is only seldom 
mentioned by historians: they were 45,000 against 26,000 Swedes and above all
they had 72 guns while the Swedes had only four (Bodart 1908, p.159). \qL

As an economic example, let us consider the almost simultaneous crashes in the
diamond, gold, palladium, platinum and silver markets at the beginning
of 1980 (Fig.1). Amazingly enough, different stories were invented in 
each of these cases to ``explain'' the crash. For the silver market, it 
was the attempt by the Hunt brothers to squeeze the market, i.e. to buy up
all available silver in order to corner short sellers (see in this 
respect Fay 1982); it should be emphasized that 
no precise data are available to support and validate that story quantitatively. 
In the diamond market,
depending upon the author, the reason invoked was the attempt by the cutters
in Tel Aviv to hoard uncut diamonds for the purpose of speculation 
or the short-lived boom in diamond investment (Boyajian 1988, Epstein 1982). 
Other stories, which we omit here for the sake of brevity, were proposed 
to account for the crashes in the other markets. It can be noted that
all these stories are (i) plausible (ii) purely qualitative 
and (iii) fairly picturesque; these features certainly explain why they
were easily memorized and used over and over again by various authors. 
Such explanations are so to say self-fulfilling rationalizations. It is
as if one would explain the fall of apples from 
from different trees in one case by a swing of the branch, in another
by the fact that the apple had become
too big, and so on, but without mentioning general causes such as the
effect of a gust of wind and the role of the gravitational attraction.

\qA{The hypothesis of a collective factor due to imitation processes}

Fig.1 suggests the existence of a common factor which may account
for the simultaneous bubbles and crashes of these markets. The hypothesis
of the existence of a common factor is a natural scientific endeavor: 
explain more data with less assumptions, in other words follow Occam's razor
principle of parsimony. The existence of a common factor would explain
for instance why the
speculative bubbles for commodities as different (in terms of demand and
supply functions) as diamonds, gold, palladium, platinum and silver burst
almost simultaneously in January 1980 as shown in figure 1.
We propose that this common factor is an imitation process between traders
and present quantitative empirical tests for this hypothesis.

\qA{Definition of a bubble}

Speculative bubbles and crashes are among the most important tumultuous
and extreme events that mark long time series of economic and financial data.
Notwithstanding their apparent intuitive meaning and their fixation in 
the minds of the public, defining and identifying bubbles and crashes is not always obvious.
With respect to crashes, a natural characterization is to define them
as outliers, i.e. cumulative drops beyond the normality (Johansen and Sornette, 1998).
Defining a bubble is even more trickly because acceptance or rejection
of the existence of a bubble is contingent on the model of fundamentals one uses:
indeed, the conclusion that a bubble exists in the data can emerge only from
using a specific model of fundamental prices (Flood and Garber, 1994). \qL

The usual conception is that a bubble can arise when the actual market price
depends positively on its own expected rate of change. In the rational expectation
framework, there is no systematic discrepancies between expectations and
realized prices. Therefore, the actual rate of 
change is positively correlated to its expected value and the actual market
price is also positively related to the actual rate of change. A simple example is
\be
{d p \over dt} = a p^b~,~~~~~~{\rm with}~b>0~.
\ee
If $b<1$, the price $p$ exhibits a power law acceleration with an upward
concavity proportional to
$(t+t_0)^{1/(1-b)}$, where $t_0$ depends on the initial price. 
If $b \geq 1$, the acceleration is so strong that the price
diverges in finite time as in a critical point (Stanley, 1987) or ``movable singularity''
(Bender and Orszag, 1978)
as $(t_c-t)^{-1/(b-1)}$, where 
the critical time $t_c$ depends on the initial price. In such conditions, the expectation 
of price changes becomes self-fulfilling and drives actual price changes
independently of market fundamentals. This is the definition of a {\it price bubble}
(Flood and Garber, 1994).

\qA{How to detect a bubble?}

Notwithstanding this deceivingly simple definition, as we said, the very notion of a bubble makes
no sense without a precise model detailing the market behavior. Indeed, without a model,
it is impossible to isolate the price trajectory characterizing a bubble: an apparent
anomalous price behavior could actually result from the rational behavior of
informed traders reacting to important news. A large literature has thus been devoted
to the specification of reasonable models and to the identification of favorable time periods,
like hyperinflation where it is believed that the work of disentangling the bubble 
trajectory from the fundamental evolution is easier (Flood and Garber, 1994). \qL

A second approach is to try to extract ``universal'' results on the expected
behavior of bubbles that are independent of the assumed underlying fundamentals.
In this respect, using insights on the behavior of
multiplicative stochastic processes and the no-arbitrage condition, 
Lux and Sornette (1999) have shown recently that rational bubbles
\`a la Blanchard and Watson (1982) have
a ÔfatÕ power tail for both the bubble price component, its price difference and its returns
with an exponent $\alpha$ {\it always} smaller than $1$. In contrast, the usual empirical estimates
give $\alpha \approx 3-4$. It therefore appears that
exogenous rational bubbles are hardly reconcilable with some of the stylized
facts of financial data at a very elementary level.  \qL

Another approach developed in the last few year using analogies with physical 
phenomena is to characterize a highly specific signature that can be 
used as a fingerprint for a bubble. A notable example is the
robust and apparently universal signature found in bubbles ending in
large crashes or large corrections found in major financial stock markets as well
as in emergent markets, namely accelerated
price increase with a sharp peak (Roehner and Sornette, 1998)
decorated by large scale log-periodic oscillations accelerating up to
 a kind of critical point (Johansen, Sornette and Ledoit, 1999; 
Johansen and Sornette, 1999 and references therein).  \qL

Here, we propose yet another method which bypasses all the difficulties
of the previous approaches by monitoring external indicators which show 
an anomalously growing interest in the public at times of bubbles. Such
indicators can be for instance the number of publications and books published on the topic
and the size of their sales during bubble periods. From the definition of a bubble
as a self-fulfilling reinforcing price change, we thus search for
indicators of a possible self-reinforcing imitation
between agents in the market. We will show that such
tendency for traders to imitate their nearest ``neighbors'' 
is self-reinforcing and increases
up to a certain point called the ``critical'' point, at which all traders may
place the same order (sell) at the same time, thus causing the burst of
the bubble (a crash, a large correction or more generally a change of regime). 
The main point of our paper is that, during a major speculative bubble, the fever does
not remain confined to the economic or financial sphere but spreads to 
other segments of the society which can actually become actors themselves
by buying the market, usually close to the end of the bullish mood. This
``fever contamination'' provides a probe of the imitation process
which is at the source of the speculative bubble, and gives therefore a direct
access to the fundamental mechanism of the bubble. We believe that this
approach is better suited to qualify the existence of bubble than other
methods relying on poorly constrained models of the fundamental prices.

\qA{Interaction channels}

How can speculative
fever spread? Sociologists distinguish three kinds of communication
channels (Deutsch 1953): (i) inter-personal communication i.e. contacts between 
individuals by word of mouth, phone, email, etc; (ii) commercial
communication channels such as newspapers, magazines, books, 
radio, television, films, etc.; (iii) institutional channels such as churches,
political parties, etc. In terms of interaction range, the channel (i) is basically
a short-range interaction in the (not necessarily spatial) sense that most
of the persons that you know personally belong to the same segment 
of society while channel (ii) and also to some extent channel (iii) are ``long range'' 
interactions. In order to assess the role played by these channels, the
crucial point is to find sources from which one can draw {\it quantitative}
evidence. In this study, we restrict ourselves to one specific
commercial channel namely the publishing business. 

\qA{Content of the paper}

In section 2, we propose a simple model of price
evolution that shows that imitation may result from rational expectations.
The resulting Ising model dynamics embodies the possible existence of
a speculative bubble and its culmination at a critical point.
In section 3,  we show how
the progress of the speculative frenzy can be estimated through the 
production of books on related topics. In section 4, we use the volume of
transactions in a given market as a measure of its widening attraction 
power.

\qI{Imitation results from rational maximization of expected profit}

\qA{Definition of the model}

In this section, we follow and generalize the work of Bhamra (1999).
Consider $N$ traders in a network, whose links represent the communication channels
through which the traders communicate. The graph is not euclidean in general
and is probably better described by the ``small-world'' type of topology (Watts, 1999)
found to describe the short chain of intermediate
acquaintances between any two people in the world (colloquially referred to as the 
``six degrees of separation'').
We denote $N(i)$ the number of traders directly connected to $i$ on the graph.
The traders buy or sell one asset at price $p(t)$ which evolves as a function of
time assumed to be discrete and measured in units
of the time step $\Delta t$. In the simplest version of the model, each agent can
either buy or sell only one unit of the asset. This is quantified by 
the buy state $s_i=+1$ or the sell state $s_i=-1$. Each agent can trade at time 
$t-1$ at the price $p(t-1)$ based on all previous informations including that at $t-1$.
We assume that the asset price variation is determined by the following equation
\be
{p(t)-p(t-1) \over p(t-1)} = \Delta t~ F\left({\sum_{i=1}^N s_i(t-1) \over N}\right) ~+~ 
\sqrt{\Delta t}~ \sigma~ \eta(t)~.  \label{jfjfjka}
\ee
$\sigma$ is the price volatility per unit time, the factor $\sqrt{\Delta t}$ represents
the time factor of the volatility and $\eta(t)$ is a white Gaussian noise with 
unit variance.  The first term in the r.h.s. of (\ref{jfjfjka})
is the systematic price drift resulting from
the possible imbalance between buyers and sellers. We assume that the function $F(x)$ 
is such that $F(0)=0$ and is
monotonically increasing with its argument: perfect balance between buyers and sellers
does not move the price; a larger (resp. smaller) number of 
buyers than sellers drive the price up (resp. down).
An often used dependence is simply 
a linear relationship $F(x) = \mu x$. Assuming that the time needed to complete a trade of
size $L$ is proportional to $L$ and that the unobservable price fluctuations obey a
diffusion process during that time, Zhang (1999) has proposed instead $F(x) \propto 
{\rm sign}(x) \sqrt{|x|}$.
The second stochastic term of the r.h.s. of (\ref{jfjfjka}) accounts for noisy sources
of price fluctuations. Taken alone, it would give the usual log-normal random walk process
(Cootner, 1967). 

\qA{Rational expectation investment strategy}

At time $t-1$, just when the price $p(t-1)$ has been announced, the trader $i$
defines his strategy $s_i(t-1)$ that he will hold from $t-1$ to $t$, thus
realizing the profit $(p(t)-p(t-1))s_i(t-1)$. To define $s_i(t-1)$, the trader
calculates his expected profit $P_E$, given the past information and his position,
and then chooses $s_i(t-1)$ such that $P_E$ is maximum. Within the rational
expectation model, all traders have full knowledge of the fundamental equation
(\ref{jfjfjka}) of their financial world. However, they cannot poll the positions
$s_j$ that will take all other traders which will determine the price drift
according to (\ref{jfjfjka}). The next best thing that trader $i$ can
do is to poll his $N(i)$ ``neighbors'' and construct his prediction for the 
price drift from this information. The trader needs an additional information, namely
the a priori probability $P_+$ and $P_-$ for each trader to buy or sell. 
The probabilities $P_+$ and $P_-$ are the only informations that he can use for all the
traders that he does not poll directly. From this, he can form 
his expectation of the price change. The simplest case corresponds to a neutral 
market where $P_+=P_-=1/2$. To allow for a simple discussion, we restrict the discussion
to the linear case $F(x) = \mu x$. The general nonlinear case complicates the matter
and a careful convexity analysis must be performed. 
In the linear case, the trader $i$ expects the following price change
\be
\Delta t~ \mu~ \left({\sum_{i=1}^{*~N(i)} s_i(t-1) \over N}\right)~+~\sqrt{\Delta t}~ \sigma~ \eta(t)~.
\ee
Notice that the sum is now restricted to the $N(i)$ neighbors of trader $i$ because
the sum over all other traders, whom he cannot poll directly, averages out. This restricted
sum is represented by the star symbol.
His expected profit is thus
\be
\left(\Delta t~ \mu~ \left({\sum_{i=1}^{*~N(i)} s_i(t-1) \over N}\right) +
\sqrt{\Delta t}~ \sigma~ \eta(t)\right)~p(t-1)~s_i(t-1)~.
\ee
The strategy that maximizes his profit is 
\be
s_i(t-1) = sign\left( {\mu \over N} \sum_{i=1}^{N(i)_*} s_i(t-1) ~+~ {\sigma \over \sqrt{\Delta t}}
~ \eta(t)\right)~. \label{jgfhha}
\ee

\qA{Stochastic dynamical model of interacting particles}

Equation (\ref{jgfhha}) is exactly the evolution equation postulated 
by Johansen, Sornette and Ledoit (1999)
and by Johansen, Ledoit and Sornette (2000)
for the dynamics of imitation between agents. This evolution equation (\ref{jgfhha})
belongs to
the class of stochastic dynamical models of interacting particles
(Liggett, 1985, 1997), which have been much studied mathematically in the
context of physics and biology.
In this model (\ref{jgfhha}), the tendency
towards imitation is governed by $\mu/N$, which is called the coupling strength;
the tendency towards idiosyncratic behavior is governed by $\sigma$. Thus
the value of $\mu/N$ relative to $\sigma$ determines the outcome of the battle
between order (imitation process) and disorder, and the development of a bubble.
More generally, the coupling strength $\mu/N$ could be heterogeneous across pairs of
neighbors, without substantially affecting the properties of the model.
Some of the $\mu_{ij}$'s could even be negative, as long as the average of all
$\mu_{ij}$'s was strictly positive.
The equation (\ref{jgfhha}) only describes the state of an agent at
a given point in time. In the next instant, new $\eta$'s are drawn, new
influences propagate themselves to neighbors, and agents can change states.
Thus, the best we can do is give a statistical description of the states.  \qL

The model does {\it not} assume instantaneous opinion
interactions between neighbours. In real markets, opinions tend indeed not to be instantaneous
but are formed over a period of time by a process involving family, friends,
colleagues, newspapers, web sites, TV stations, etc.
Decisions about trading activity of a given agent may occur when the consensus
from all these sources reaches a trigger level. This is precisely this feature
of a threshold reached by a consensus that expression (\ref{jgfhha}) captures:
the consensus is described by the sum over the $N(i)$ agents connected to agent $i$
and the threshold is provided by the sign function together with the idiosyncratic
signal included in $\eta$. The delay in the formation of the opinion of a given 
trader as a function of other traders'opinion is captured
in our model by the progressive spreading of information 
during successive updating steps (see for instance (Liggett, 1985, 1997)). \qL

The simplest possible network is a
two-dimensional grid in the Euclidean plane. Each agent has four nearest
neighbors: one to the North, one
to the South, the East and the West. The relevant parameter is
$K\equiv \mu /\sigma$. It measures the tendency towards imitation relative to
the tendency towards idiosyncratic behavior. In the context of the alignment
of atomic spins to create magnetisation, this model is identical to the
so-called two-dimensional Ising model which has been solved explicitly by Onsager
(1944). Only its formulation is different from what is usually found in textbooks 
(Goldenfeld, 1992), as we emphasize a dynamical view point. \qL

In the Ising model, there exists a critical point $K_c$ that determines the
properties of the system. When $K<K_c$, disorder reigns: the
sensitivity to a small global influence is small, the clusters of agents
who are in agreement remain of small size, and imitation only propagates between
close neighbors. Formally, in this case, the susceptibility $\chi$ of the
system is finite. When $K$ increases and gets close to $K_c$, order
starts to appear: the system becomes extremely sensitive to a small global
perturbation, agents who agree with each other form large clusters, and
imitation propagates over long distances. In the Natural Sciences, these are
the characteristics of so-called {\em critical} phenomena. Formally, in this
case the susceptibility $\chi$ of the system goes to infinity.  The
hallmark of
criticality is the {\em power law}, and indeed the susceptibility goes to
infinity according to a power law:
\be
\label{eq:power}
\chi\approx A(K_c-K)^{-\gamma}.
\ee
where $A$ is a positive constant and $\gamma>0$ is called the {\em critical
exponent} of the susceptibility (equal to $7/4$ for the 2-d Ising model).
This kind of critical behavior is found in many other models of interacting elements
(Liggett, 1985, 1997). In the case of a network where all traders are connected
to all other traders, one obtains the so-called ``mean-field'' critical behavior,
for which $\gamma=1$. \qL

We view a bubble as resulting from imitation processes similar to that described
in this class of models, in which the ``coupling'' strength $K$ is larger than the
critical value so that a globally bullish mood is established {\it independently}
of the fundamentals. Another scenario is that $K$ itself evolves as a function of time, 
progressively increasing up and beyond the critical value $K_c$. Both scenarii
lead to a bubble, i.e. to a price increase which is decoupled from the fundamentals.

\qI{The production of books reflects the spread of the bubble} 

In this section, we use the number of books published on a given topic as
an observational probe of the mood and crazes of the
society at a given moment; more specifically, we list the number of 
books on a given subject in the catalogues of major libraries in 
order to assess to what extent this issue is a matter of concern or 
interest. As will turn out, this measure provides a surprisingly good
``thermometer''. The use of the terminoly ``thermometer'' is not just
a way of speaking. The model presented in section 2 actually defines the
temperature as $1/K$ in proper units. As we have seen, this parameter controls
the strength of imitation between agents, which in this class of models is indeed
controlled by the temperature.
Before we use this thermometer for our purpose,
it is appropriate to test it by performing a preliminary experiment. 

\qA{Preliminary test: assessing inflation from the number of published books}

Let us estimate the number of books published about inflation every
year. To this aim we use the catalogues of two university
libraries, namely: (i) The Harvard library which is a major general   
library; (ii) The library of the National Foundation for Political Sciences
(FNSP, Paris), which is a library specialized in economics and political 
sciences. Furthermore, in order to test whether the results depend upon 
the counting procedure, we used two different criteria. At Harvard, we
counted all the books whose titles contain the word ``inflation''; at the
FNSP, we counted the books listed under the ``inflation'' heading of the
subject classification; in spite of the fact that
this second criterion depends to some extent on
the judgment of the librarian in charge of the subject index,
it will be seen that both criteria lead to similar results. 
Fig.2 shows the two curves for the number of books in each library along
with the curve of the inflation rate in the United States (dotted line).
There is obviously a close connection between the three curves; the three
cross-correlations are over 0.82. For instance, the correlation between the
Harvard book data and the inflation rate is 0.89; the correlation between the
Harvard data and the FNSP data is 0.87. \qL

These observations lead to the following conclusions: (i) when inflation
becomes higher, the concern about price increase seems to pervade large 
sections of the society: it makes publishers more prone to publish books
on that topic, authors more willing to write them, librarians more ready to
buy them and the public more desirous to read them. 
In fact the correlation is so good that the level 
of inflation could be measured fairly accurately  
(albeit with a time lag of one or two years) through the number of books
published on that topic. 
(ii) Estimates of the
number of books published on an issue of worldwide interest are fairly 
robust with respect to the country where the library is located, the type
of the library (general versus specialized for instance) or the method used
(e.g. whether based on a title or subject search). \qL

In order to emphasize the significance of the previous result, the
following comparison may be helpful. When a new experiment in the 
field of particle physics is carried out, one would certainly expect
an increased production of papers on that topic. What we have observed
in the case of inflation is fairly different however for at
least two reasons. (i) The book/inflation relationship is not restricted
to major inflation peaks; even modest upsurges in times of low inflation
(under 5\%) brought about an increased book production. (ii) The study 
of the inflation phenomenon is by no means facilitated in times of 
high inflation. On the contrary, from a scientific point of view, one
would be in a much better position 5 or 10 years after an inflation
peak for one could then study both the upward and downward phase of the
inflation episode. \qL

In short, Fig.2 shows that the society reacted with great sensitivity
(so to say as a resonance chamber) to fluctuations of the price
index. It is important to note that, in this case, one cannot argue that
the increased book production stimulated inflation. On the contrary, 
in the case of speculative trading to which we turn in the next section,
an augmented book production has a positive feedback effect on the 
economic phenomenon it addresses: it feeds speculation by sustaining the public's interest.

\qA{Connection between speculative trading and production of books}

In this paragraph, we propose three examples which point to a close connection
between a speculative bubble in a given sector and the production of new
books on that topic. All these experiments have been conducted on the
electronic catalogues of the Harvard Library; it can 
be accessed through the instruction: telnet hollis.harvard.edu. \qL

The first curve (solid line) in Fig.3a shows the number of books 
published yearly whose titles contain one of the words: ``stocks'', 
``stock market'' or ``speculation''. The second curve (broken line) gives
the level of stock prices. 
The connection between both phenomena is apparent especially during the 
peak of 1925-1932. The book production lags behind the price index, the time
lag being of the order of 1.5 years,
which is approximately the time it takes to 
write and publish a book; with this time lag taken into account, the correlation
is 0.57; of course it would be substantially higher if the series 
were restricted to the 1925-1932 peak. \qL

Fig.3b tells a similar story for a more recent time period, namely the
1980-1996 bull market. The correlation is 0.70 (a zero- or a one 
year-lag lead almost to the same correlation). \qL

The last example concerns the diamond bubble of 1975-1984
(Fig.3c). There is a close
connection between the number of books whose titles contain the words
``diamond'' or ``diamonds'' and the actual price of diamonds. The correlation
(without lag) is 0.66. \qL

The augmented book production is of course but one manifestation of the 
society's increased interest for speculative trading. By its very nature,
because of the time it takes to write and publish a book, this indicator 
displays considerable inertia. Needless to say it would be of interest to
find an indicator responding more quickly to a change in the society's mood. 
The next paragraph provides such an example. 

\qA{The coin bubble of 1965}

This example is taken from a very stimulating study by Montroll and Badger 
(1974, p.200). In the early 1960s, there was a speculative bubble for
a number of American coins. For instance, a roll of fifty 1960 Denver-minted
pennies of the small date variety which sold for 4 dollars in 1961 fetched 
21 dollars in 1964; similarly a roll of 1955 half dollars went from 
20 dollars to 190 dollars in that period. One of the main journals in the
field of coin collecting is ``Coin World'' which has appeared weekly since
1959. Undoubtedly a weekly publication can reflect the speculative mood of the
market more swiftly and accurately than newly published books. The solid
line curve in Fig.4 shows the number of paid copies of ``Coin World'' that have been
distributed. The curve began to level off by October 1964 and the maximum 
was reached on March 1, 1965. On the same figure, we plot the price
history of the 1960\ D small date penny and of the 1955 half dollar; these curves
are smoothed least square fits to the weekly price fluctuations 
given in ``Coin World''. \qL

Once again there is a conspicuous parallelism and this time the distribution
curve does not lag behind the price series; the leveling off of distribution
even precedes the turning point of the half dollar market. In this case,
the distribution data could have been used as a warning signal. 
Of course, for the penny price, the leveling off occurred even earlier,
but the price of one coin can hardly be used to predict the price of another
coin; in contrast the fluctuations in ``Coin World'' distribution are
a counter of the number of people who have a general interest in the coin
market. Note that the sharp peak versus flat trough asymmetry 
(Roehner and Sornette, 1998) have been found in a simple percolation model 
of the stock market (Stauffer and Sornette, 1999) by 
letting the trading activity be dependent on the price,i.e.
increasing prices
causes more people to act than a decreasing price.  \qL

It is of interest to consider for a
moment actual price increases. The price of the 1960\ D penny increased
by a factor 5.2 (from 4 to 21 dollars), while the price of the 1955
half dollar was multiplied by 9.5 (from 20 to 120 dollars). This provides
a new confirmation of the price multiplier effect (Roehner 2000) which 
says that for a more expensive item  the speculation will be stronger 
and lead to a larger price increase. \qL

What is the rationale of the effect described in this section or,
in other words, why does the interest shown by the public for a given item
reflect actual price changes during a speculative bubble? Basically each
new person showing an interest in the market is a potential customer. 
When the number of interested people levels off, no new customer appears. 
Observation shows that the high level
attained by the market can be sustained
only for a short time with ``old customers'' buying from and selling to
each other. Theoretically, these customers could also 
decide to devote a larger
share of their revenue to buying coins; in this case, the bubble could 
continue to grow even with a fixed pool of customers. However, observation
shows that this does {\it not} happen, at least in the cases that we 
discussed. It can be added that for an item as the 1960\ D penny that costs
only 20 dollars even at its peak price, it would be fairly easy for 
many collectors to buy more; 
nevertheless this does not happen: as soon as the
number of new customers entering the market begins to level off, the 
bubble begins to falter. This very same mechanism has been documented by Galbraith (1997)
in his famous analysis of the 1929 crash and speculative bubble preceeding it.
More recently, Krugman (1994, 1995, 1998) has shown that the same mechanism has been at work in the
speculative bubble preceeding the Asian collapse a few years ago.
\qL

Can the same
investigation procedure be applied to the stock market? The most serious
difficulty is to find a market journal which is representative of the level 
of public interest. The ``Wall Street Journal'' has been the most important 
newspaper in the securities field for many years; however, it has also
achieved recognition as a national newspaper so that it is read for general 
information as well as for investment news. As a matter of fact, its average
yearly circulation does not closely follow the fluctuation of stock prices. 
For instance, during the bear market of 1968-1974, its circulation 
{\it increased}
from 1.1 million copies to 1.3 million, i.e. a 18\% increase. 
That increase was however much smaller than the one that occurred 
during the 1960-1968 bull market, namely 72\%. In short, the Wall Street
Journal to some extent shows the phenomenon described above but because of
its broad spectrum  it cannot serve as an accurate thermometer.

\qI{Volume of trade as an indicator of the public's interest}

For stock markets, there is a vast literature on the (possible) relationship
between trading volume and prices: e.g. Crouch (1970),
Rogalski (1978), Schneller (1978), Karpoff (1987); Crouch's contribution
is particularly interesting. However, most of these papers are
concerned with short-term (daily or weekly) changes; since both prices
and volumes are fairly erratic (the latter being even more erratic than
the former), this turns out to be a difficult statistical question. 
In this section, we are concerned with long term (yearly) variations. 
Moreover, we use the volume of sales as an indicator of the public's interest
for a given speculative item; in that sense, the present section is a
logical continuation of our analysis in the previous section. 

\qA{The stock market}

Wall Street has several adages about trading volumes, for instance:
``It takes volumes to make prices  move'' (cited in Schneller 1978) or
``Transaction volume [i.e. the number of shares traded] tends to 
be high in bull markets and low in bear markets'' (cited in Karpoff
1987, p.112). More precisely it will be seen that transaction volumes
are high {\it and growing} in bull markets and low {\it and decreasing}
in bear markets. Moreover, this conclusion holds not only for stock
markets but also for other speculative markets, for instance the land and
property markets. \qL

The volume of shares traded on the NYSE between 1895 and 1940 is 
plotted in Fig.5a (solid line) along with the (deflated) Standard and
Poor stock price index. Between 1895 ad 1914, there is a clear parallelism
between volume and price (the correlation is 0.60). During the bull market
of the 1920s, the yearly volume of shares traded increased from 0.2 billion
to one billion; this five fold increase matched a multiplication
by three of the price index (the correlation is 0.76). Similarly,
Fig.5b shows that, during the bull market of the 1980s and 1990s, there was 
a graphic connection between trading volume and share prices: the volume
increased from about 10 billions to more than 100 billions, a ten fold increase
which matches an eightfold increase of the price index (the correlation is
0.98). The huge increase in the volume of transactions is a direct proof of the
fact that there has been a permanent inflow of new customers into the market;
that inflow is itself a manifestation of a widening public interest for 
stock investment. In this case, this effect was probably accompanied by an 
increase of the funds each customer was prepared to invest in the stock market. 
This is in particular suggested by the series given in the following table. 
\vskip 4cm

\centerline{\bf Table 1 \ Distribution of transactions by size (NYSE)}

\def\tvi{\vrule height 12pt depth 5pt width 0pt}

$$ \matrix{
\tvi &    & 1975 & 1976 & 1977 & 1978 & 1979 & 1980 & 1981 & 1982 & 1983 \cr
\noalign{\hrule}
\tvi \hbox{100-4900 shares} \hfill & (\%) 
& 73.5 & 70.2 & 66.0 & 64.1 & 61.2 & 56.9 & 52.1 &
 44.7 & 41.3 \cr
> \hbox{5000 shares} \hfill & (\%) 
& 26.6 & 29.8 & 34.0 & 35.9 & 38.8 & 43.1 & 47.9 &
 55.3 & 58.7 \cr
 & & & & & & & & & & \cr
\tvi  &   & 1984 & 1985 & 1986 & 1987 & 1988 & 1989 & 1990 & 1991 & 1992  \cr
\noalign{\hrule}
\tvi \hbox{100-4900 shares} \hfill & (\%) 
& 36.4 & 34.7 & 36.5 & 34.7 & 32.4 & 34.1 & 35.2 &
 34.8 & 33.5 \cr
> \hbox{5000 shares} \hfill & (\%)
& 63.6 & 65.3 & 63.5 & 65.3 & 67.6 & 65.9 & 64.8 &
 65.2 & 66.5 \cr
} $$
{\it Source: Statistical Abstract of the United States (various years)} 
\vskip 4mm

Thus, the number of transactions of more than 5,000 shares (i.e. more than 
50,000 dollars if one takes the price of a share to be of 
the order of 10 dollars) has increased steadily from 1975 on, which was the
first year for which this statistics was given. But the rate of growth which 
has been very rapid between 1975 and 1984 has become much slower afterwards. 
It would be of great interest to know if the progression has continued after
1992, unfortunately these data are no longer given after this date. 

\qA{Real estate markets}

When turning from stock markets to real estate markets, one is confronted
with two difficulties. The first is to identify a genuine ``bull market'',
the second is to find adequate statistics, which is not easy especially  
for the number of 
transactions. Fig.6a and b provide two examples. The first one concerns sales
of land in Paris; the solid line shows the number of transactions, while
the broken line gives the price per square meter; during the 1982-1991 bull 
market, both curves were parallel
(correlation is 0.95). The second example concerns the sales of apartments
in Paris shown in Fig.6b. The solid line shows the volume of sales in thousands of 
square meters, while the broken line shows the price per square meter. 
The story is somewhat different for this case. During the bull market phase,
sales were at a high level but experienced only a slow increase; then
in 1987, almost three years before the turning point for the prices, sales
began to plummet. The volume of sales could have be used as a useful indicator
of an incoming change of regime similarly to the case of the distribution of
World Coin preceding the drop of the 1960 D penny. What is not obvious is to 
estimate a priori the time lag, i.e. how low the high price of apartments 
could be sustained in the presence of decreasing demand.

\qI{Summary}

This paper has proposed new indicators for identifying speculative bubbles,
defined as self-fulfilling reinforcing price changes.
This work started with the recognition that qualifying the existence of speculative
bubbles is controversial due in part to the need for a model of fundamental prices.
Tests of the existence of a bubble then also become tests of the model
and rejection cannot be attributed uniquely to the absence of a bubble but is linked
to the quality of the model. In this standard approach, there is no unique answer.
Here, we have presented a novel empirical method which bypasses these difficulties
by monitoring external indicators of
an anomalously growing interest in the public at times of bubbles. 
We have shown the existence, during the 
growth of the bubble, of a growing interest in the public for the 
commodity in question, whether it consists in stocks, diamonds or coins. 
This interest can be quantified for instance by the increase in
the number of books published on the topic or in the increase in the subscriptions
to specialized journals. 
We have also presented a simple model
of rational expectation which maps exactly onto the Ising model on a random graph.
This model allows us to view a bubble as a result of the fight between disorder 
(idiosyncratic signal which may be different for each trader) and order (resulting
from an imitation or influence process that tend to align the opinions or interests
of people). The bubble is found to be a regime of low ``disorder'', i.e. low
``temperature'' for which the imitation processes are strong. The indicators
can then be interpreted as ``thermometers'' of the system. This model provides
a theoretical basis for spin-like models of influences discussed earlier in this context.

{\bf Acknowledgment} \ We are grateful to Ms Claire de Buttet from the
Paris bureau of the Wall Street Journal for kindly sending us the journal's circulation data.
DS acknowledges stimulating discussions with A. Johansen.

\vfill \eject

\centerline{\bf \Large References}

\vskip 1cm

\qparr
ANDERSSON (I.) 1973: Histoire de la Su\` ede des origines \` a nos jours. 
Editions Horvath. Paris. 

\qparr
BENDER (C.), ORSZAG (S.A.) 1978:
Advanced Mathematical Methods for Scientists and Engineers. McGraw-Hill, New York.

\qparr
BHAMRA (H.S.) 1999: Imitation in financial markets.
Working paper.

\qparr
BLANCHARD (O.J.), WATSON (M.W.) 1982: Bubbles, Rational Expectations and
Speculative Markets, in: Wachtel, P. ,eds., Crisis in Economic and Financial
Structure: Bubbles, Bursts, and Shocks. Lexington Books: Lexington.

\qparr
BODART (G.) 1908: Milit\" ar-historisches Kriegs-Lexikon (1618-1905). 
C.W. Stern. Vienna. 

\qparr
BOYAJIAN (W.E.) 1988: An economic review of the past decade in diamonds.
Gems and Gemology 24,134-153.

\qparr
COOTNER (P.H.) ed. 1967:
The random character of stock market prices. 
Cambridge, Mass., M.I.T. Press.

\qparr
CROUCH (R.L.) 1970: The volumes of transactions and price changes on the
New York Stock Exchange. Financial Analysts Journal 26,4,104-109. 

\qparr
DEUTSCH (K.W.) 1953: Nationalism and social communication. MIT Press.
Cambridge (Ma). 

\qparr
EPSTEIN (E.J.) 1982: The diamond invention. An expos\' e of the international
diamond monopoly. Hutchinson. London. 

\qparr
FAY (S.) 1982: The great silver bubble. Hodder and Stoughton. London. 

\qparr
FLOOD (R.P.), GARBER (P.M.) 1994: Speculative bubbles, speculative attacks, and
policy switching. The MIT Press, Cambridge, Massachussets.

\qparr
GALBRAITH (J.K.) 1997:
The great crash, 1929. Boston : Houghton Mifflin Co.

\qparr
GOLDENFELD (N.) 1992:
Lectures on phase transitions and the renormalization group.
Addison-Wesley Publishing Company, Reading, Massachussets.

\qparr
JOHANSEN (A.), SORNETTE (D.) 1998:
Stock market crashes are outliers. European Physical Journal B 1, 141-143.

\qparr
JOHANSEN (A.), SORNETTE (D.) 1999: Financial ``antibubbles'': 
log-periodicity in gold and Nikkei collapses. International Journal of 
Modern Physics C 10,4,1-13.

\qparr
JOHANSEN (A.), SORNETTE (D.), LEDOIT (O.) 1999:
 Financial Crashes using discrete scale invariance.
Journal of Risk 1 (4), 5-32.

\qparr
JOHANSEN (A.), LEDOIT (O.), SORNETTE (D.), 2000:
Crashes as critical points. Int. J. Theor. Applied Finance 3, No 1 (January issue).

\qparr
JOHANSEN (A.), SORNETTE (D.) 1999:
Log-periodic power law bubbles in Latin-American and Asian markets
and correlated anti-bubbles in Western stock markets: An empirical study.
submitted to the Journal of Empirical Finance  (http://xxx.lanl.gov/abs/cond-mat/9907270

\qparr
KARPOFF (J.M.) 1897: The relation between price changes and trading
volume: a survey.
Journal of Financial and Quantitative Analysis 22, 1, 109-126.

\qparr
KRUGMAN (P.) 1994:    
The myth of Asia's miracle.
Foreign Affairs 73, n6 (Nov/Dec): 62-78.

\qparr
KRUGMAN (P.) 1995:
Dutch tulips and emerging markets.
Foreign Affairs 74, n4 (Jul/Aug): 28-44.

\qparr
KRUGMAN (P.) 1998:
Asia:  What went wrong.
Fortune 137, n4 (Mar 2):32-34.

\qparr
LIESNER (T.) 1989: One hundred years of economic statistics. Facts on File.
New York. 

\qparr
LIGGETT (T.M.) 1985:
Interacting particle systems. New York : Springer-Verlag.

\qparr
LIGGETT (T.M.) 1997:
Stochastic models of interacting systems.
The Annals of Probability 25, 1-29.

\qparr
LUX (T.), SORNETTE (D.) 1999:
On Rational Bubbles and Fat Tails, submitted to the Journal of Monetary Economics.
(http://xxx.lanl.gov/abs/cond-mat/9910141)

\qparr
MONTROLL (E.W.), BADGER (W.W.) 1974: Introduction to quantitative aspects
of social phenomena. Gordon and Breach. New York. 

\qparr
ONSAGER (L.) 1944:
Physics Review 65, 117.

\qparr
ROEHNER (B.M.) 1995: Theory of markets. Trade and space-time patterns of
price fluctuations. A study in analytical economics. Springer-Verlag. 
Berlin. 

\qparr
ROEHNER (B.M.) 2000: Speculative trading: the price multiplier effect.
European Physical Journal B (to appear). 

\qparr
ROEHNER (B.M.), SORNETTE (D.) 1998:
The sharp peak-flat trough pattern and critical speculation.
European Physical Journal B 4, 387-399.

\qparr
ROGALSKI (R.J.) 1978: The relation between price changes and trading
volumes: a survey. Journal of Financial and Quantitative 
Analysis 22,1,109-126.

\qparr
SCHNELLER (M.I.) 1978: Security price changes and transaction volumes.
Comment. American Economic Review 68,4,696-697.

\qparr
SCOTT (F.D.) 1977: Sweden. The nation's history. University of Minnesota
Press. Minneapolis. 

\qparr
SORNETTE (D.), JOHANSEN (A.) 1997: Large financial crashes. Physica A 245,411.

\qparr
STANLEY (H.E.) 1987: 
Introduction to Phase Transitions and Critical Phenomena. Oxford
University Press, New York.

\qparr
STAUFFER (D.), SORNETTE (D.) 1999:
Self-Organized Percolation Model for Stock Market Fluctuations.
Physica A 271, 496-506.

\qparr
WATTS (D.J.) 1999:
Small worlds: the dynamics of networks between order and randomness.
Princeton, N.J.: Princeton University Press.

\qparr
ZHANG (Y.C.) 1999:
Toward a theory of marginally efficient markets.
Physica A 269, 30-44.

\vfill \eject

\centerline{\bf Figure captions}
\qpar

{\bf Fig.1 \quad Price bubble for diamonds, gold, platinum
and silver.} For gold, silver and platinum, the prices are in dollar per ounce.
For diamonds, the price is for a one-carat G flawless diamond. 
The collapse occurred almost simultaneously
in the four markets. Prices of palladium
and cobalt showed a similar evolution, we did not
represent them for the sake of clarity. 
{\it Sources: Journal des Finances (26 October 1978); The Economist (5 April 
1980); Diamond 1988: Special Report of the Economist}

{\bf Fig.2 \quad Comparison between the number of yearly published books
about inflation and the inflation rate.} Thick solid line: books at
Harvard library whose titles contain the word ``inflation''. Thin solid line:
number of books under the ``inflation'' heading of a subject classification 
(library of the National Foundation for Political Science, Paris). Dotted
line: Consumer price index in the United States. The inflation curve has
been shifted by one year to the right. The correlation between 
the number of books published and the (shifted) inflation series
is comprised between 0.82 (FNSP) and 0.89 (Harvard).
{\it Sources: Liesner (1989); Roehner (1995, p.381)}.

{\bf Fig.3a \quad  Comparison between the number of yearly published books
about stock market speculation and the level of stock prices (1911-1940).}
Solid line: books at
Harvard library whose titles contain one of the words ``stocks'', 
``stock market'' or ``speculation''; 
broken line: Standard and Poor index of common stocks. 
The curve of published books lags 
behind the price curve with a time-lag of about 1.5 years. 
{\it Source: The stock price index is taken from the Historical Abstract
of the United States.}

{\bf Fig.3b \quad Comparison between the number of yearly published books
about stock market speculation and the level of stock prices (1971-1996).}
Solid line: books at
Harvard library whose titles contain one of the words ``stocks'', 
``stock market'' or ``speculation'';
broken line: Dow Jones index of industrial shares.
{\it Source: The stock price index is taken from Quid (1997).} 
 
{\bf Fig.3c \quad Comparison between the number of yearly published books
about diamonds and the price of diamonds (1975-1984).}
Solid line: books at
Harvard library whose titles contain one of the words ``diamond'' or 
``diamonds''. 
{\it Source: The diamond prices are taken from 
Diamond 1988 (Special Report of the Economist)} 

{\bf Fig.4 \quad Comparison between the distribution of ``Coin World'' and
the price of two American coins (1962-1970).} The curve labeled ``Penny''
refers to the price of a roll of 1960\ D small date pennies; the curve 
labeled ``Half dollar'' refers to the price of a roll of 1955 half dollars. 
{\it Source: Montroll and Badger (1974)}

{\bf Fig.5a \quad Comparison between the number of shares traded on the NYSE and
the level of stock prices (1897-1940).} Solid line: number of shares traded;
broken line: deflated Standard and Poor's index of common stocks. 
{\it Source: Historical Statistics of the United States}

{\bf Fig.5b \quad Comparison between the number of shares traded on the NYSE and
the level of stock prices (1980-1997).} Solid line: number of shares traded;
broken line: Dow Jones index (industrials). 
{\it Source: Statistical Abstract of the United States, various years}

{\bf Fig.6a \quad Comparison between the number of transactions and 
the price of land in Paris.} Solid line: number of transactions; broken line:
price per square meter in thousands of French francs.
{\it Source: Le march\' e immobilier fran\c cais. Les chiffres et les
sources (1993).}

{\bf Fig.6b \quad Comparison between sales and prices of apartments in Paris.} Solid line: 
Sales in thousand of square meters; broken line:
price of apartments per square meter in thousands of French francs (of 1990).
These data refer to ``old'' apartments i.e. apartments which have
already been sold at least once. 
{\it Source: Le march\' e immobilier fran\c cais. Les chiffres et les
sources (1998).}

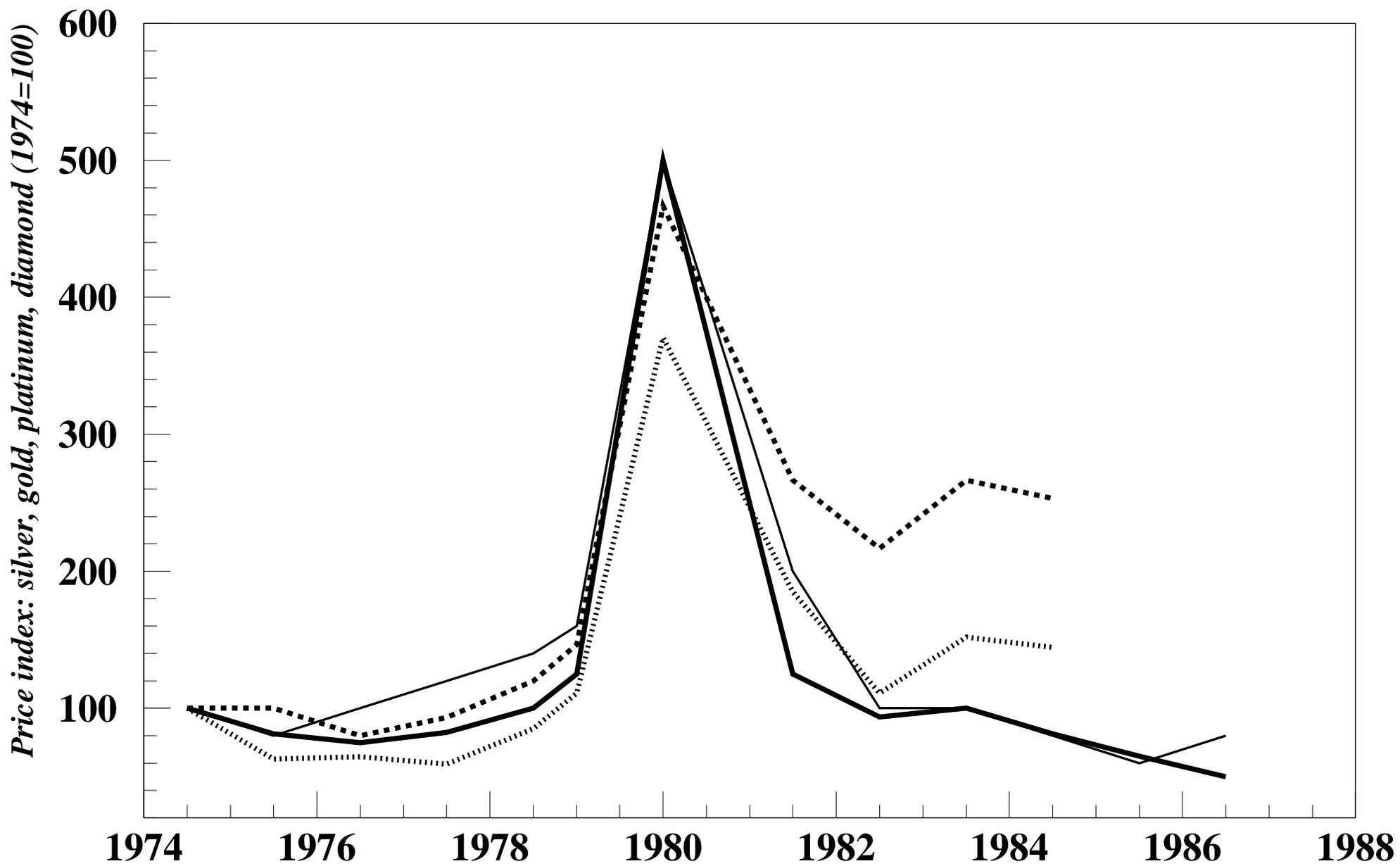
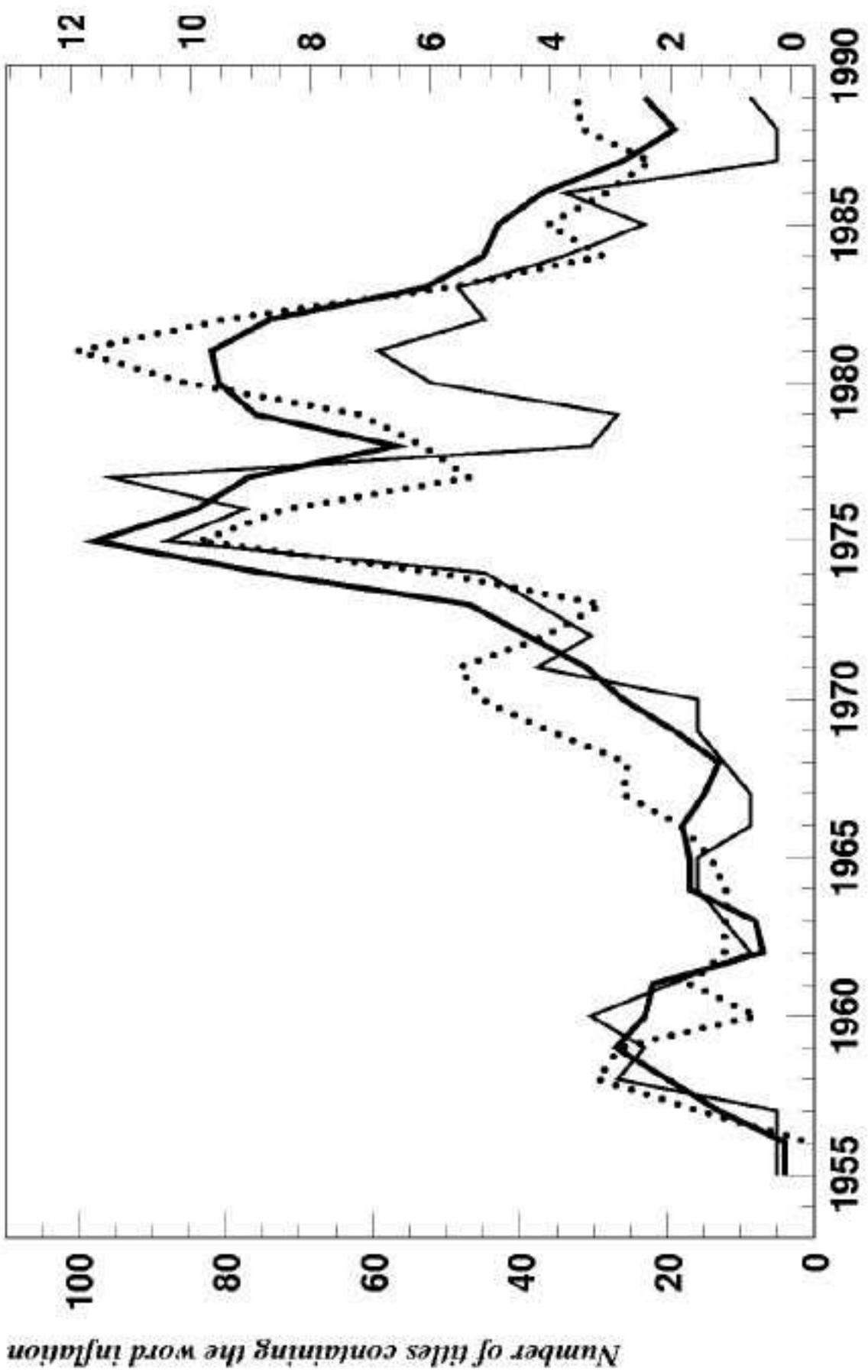
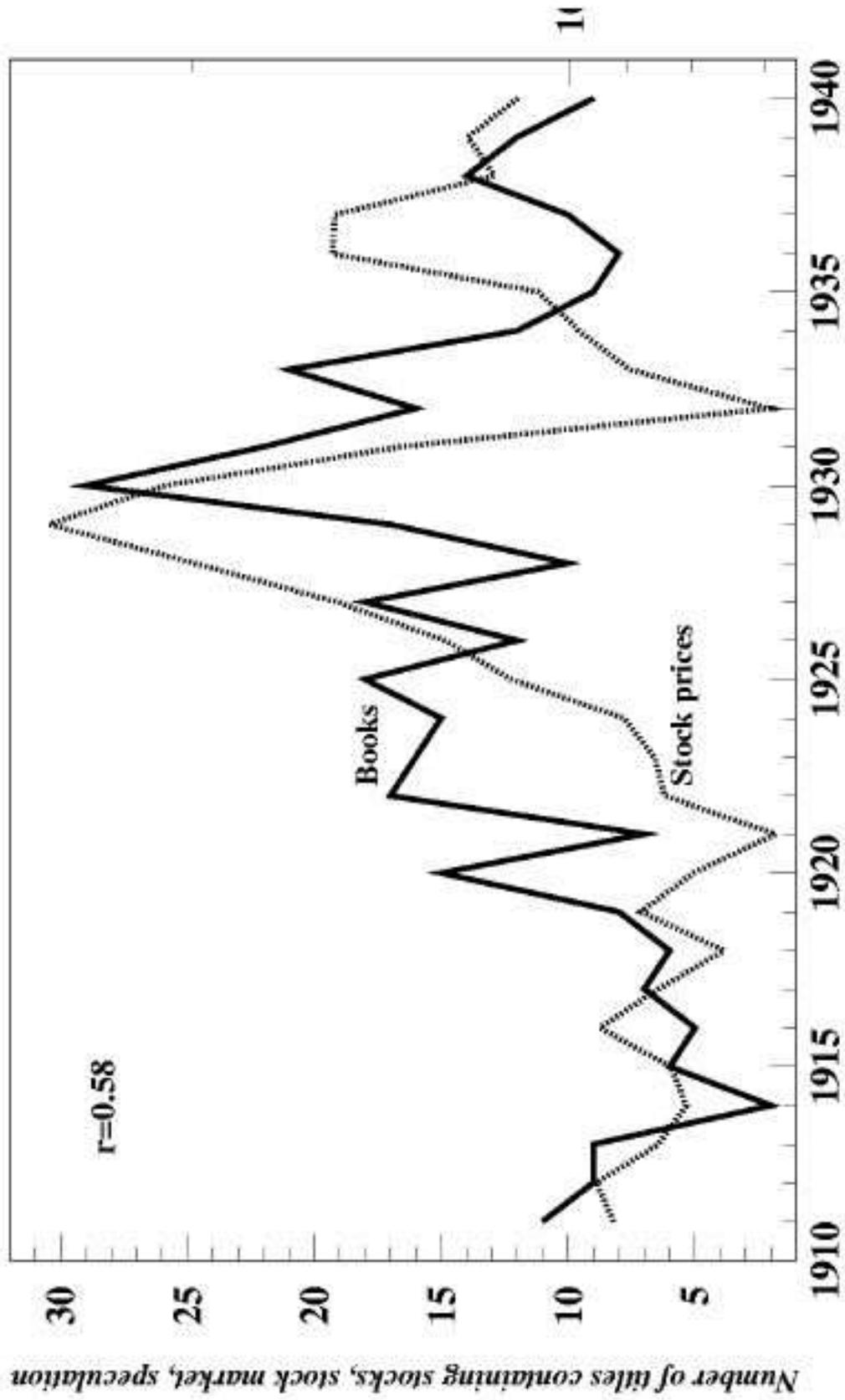
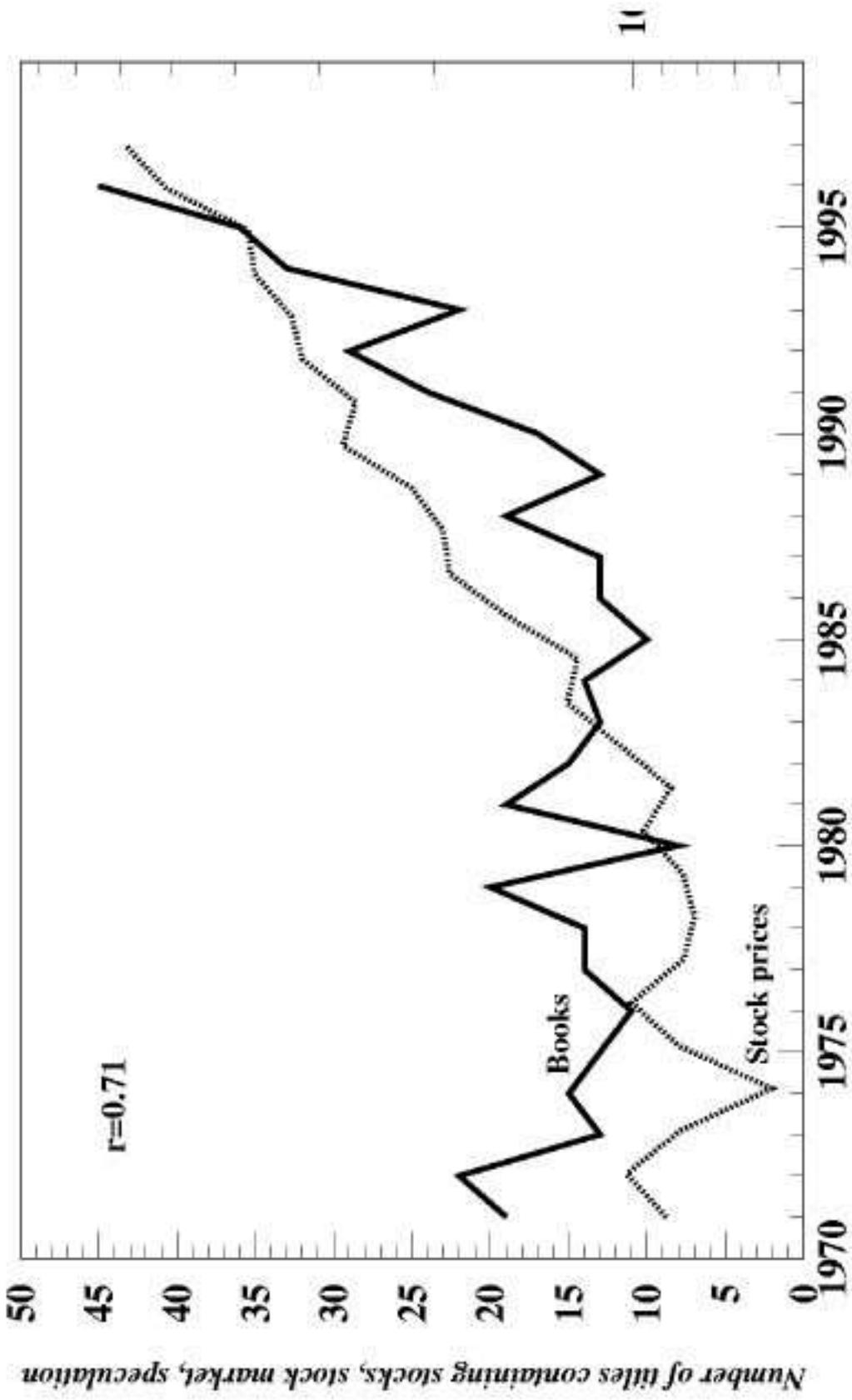
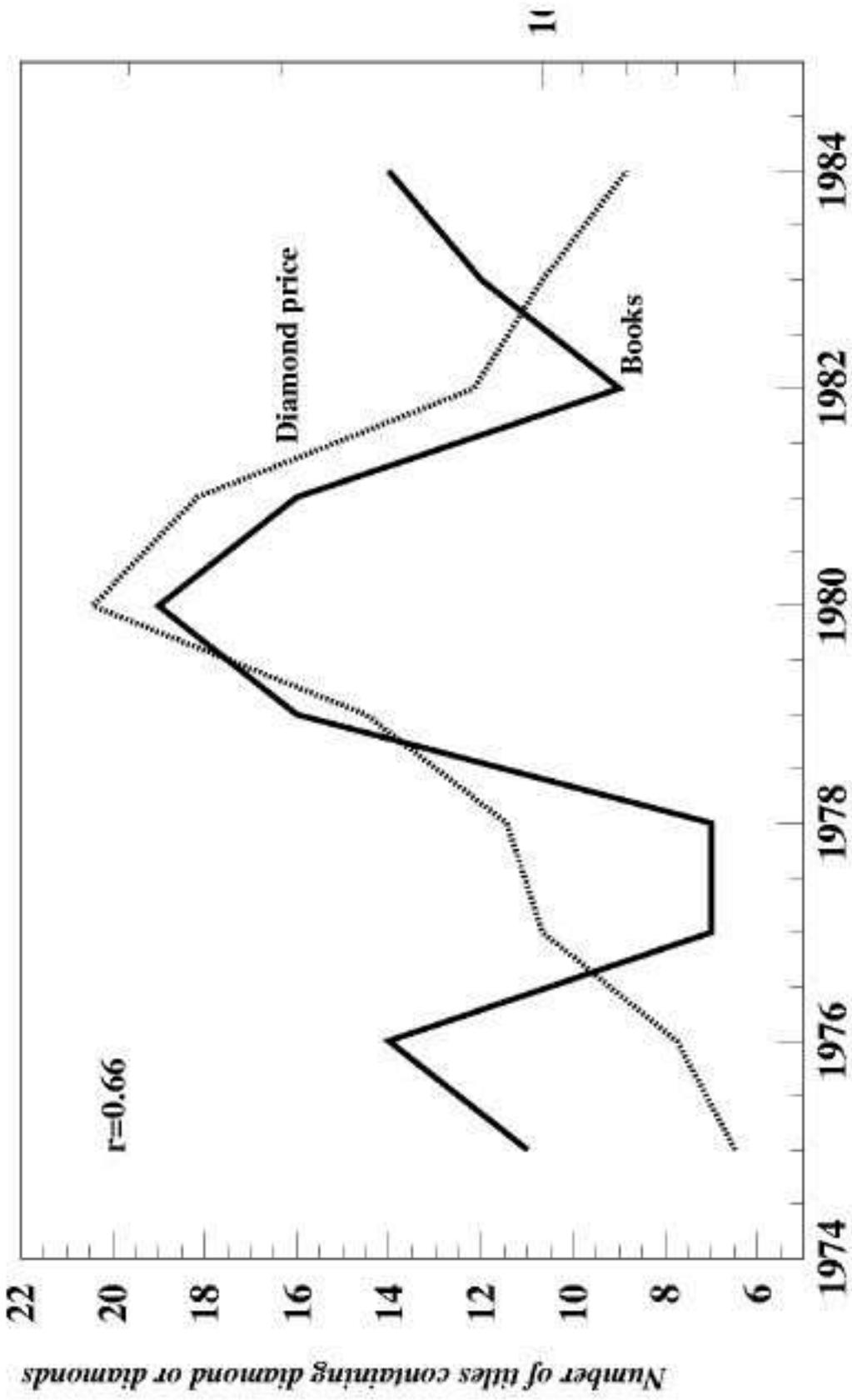
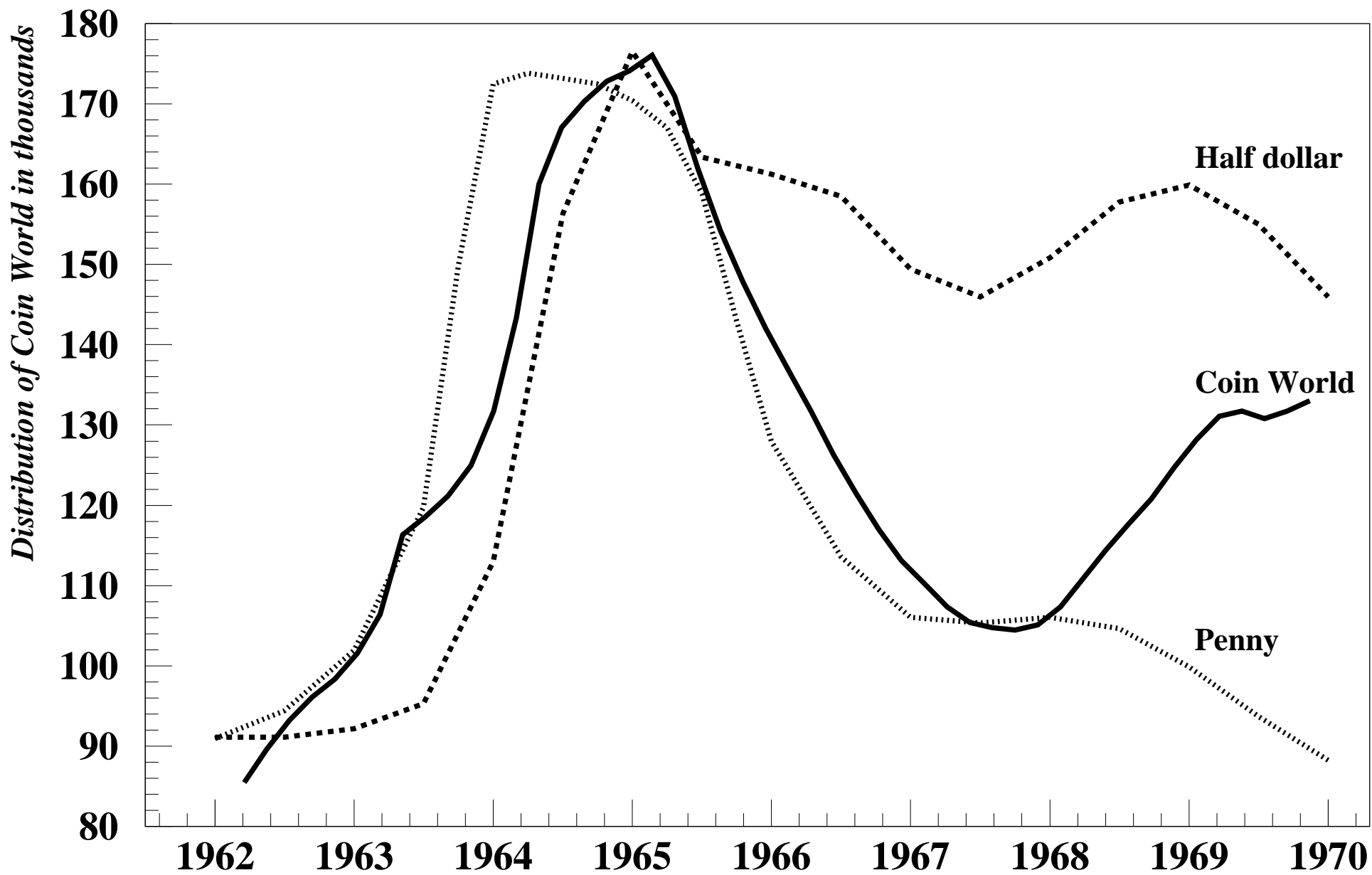
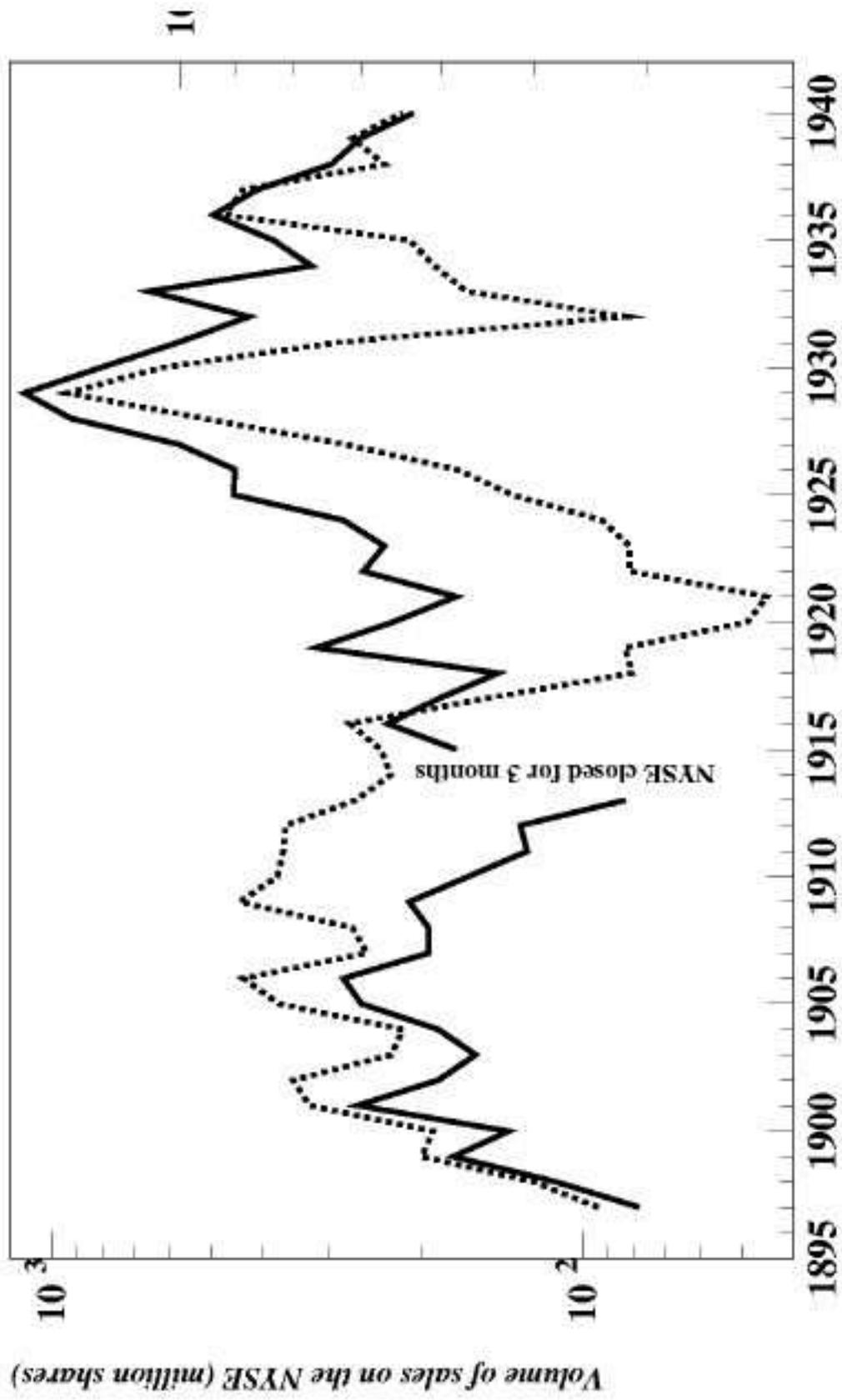
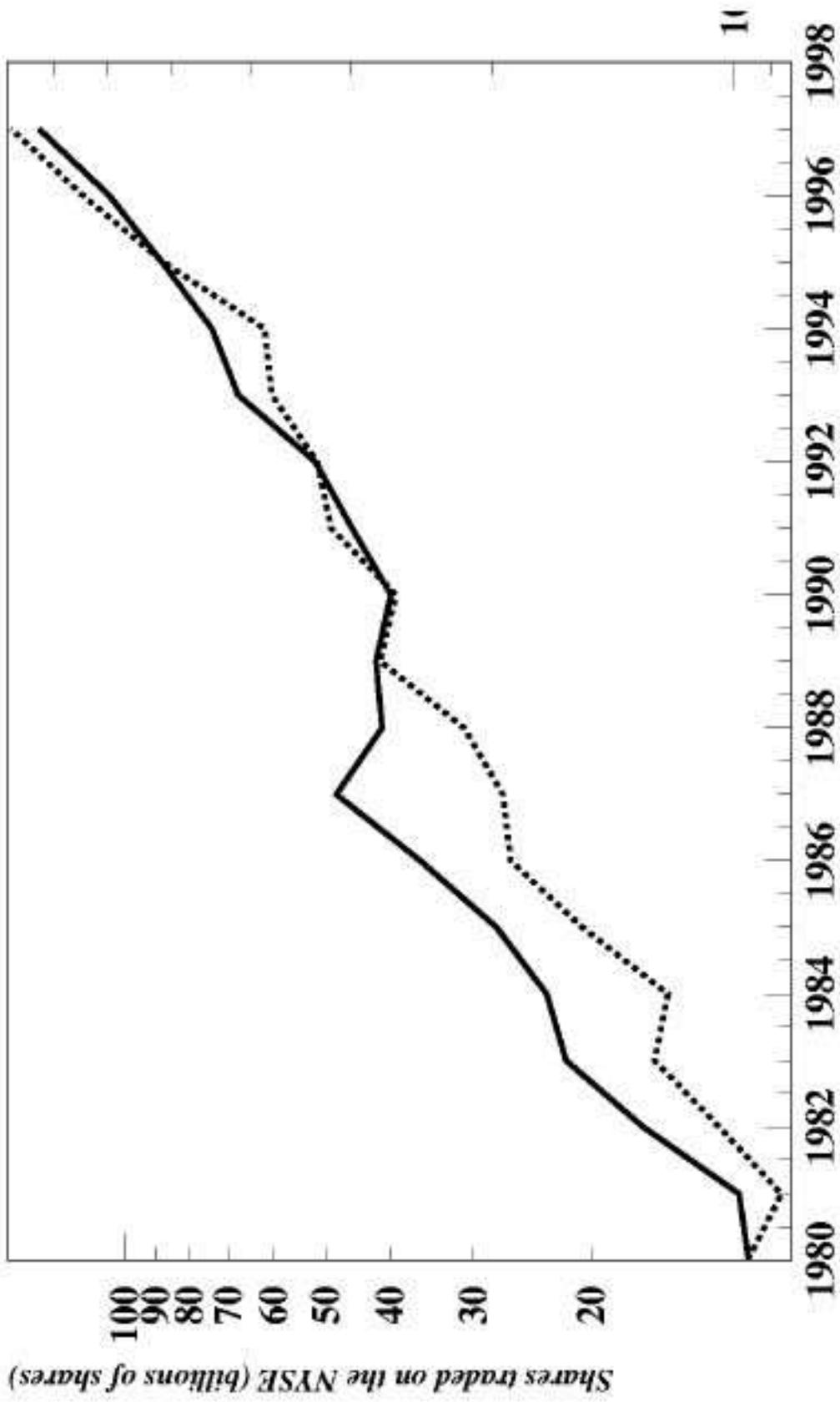
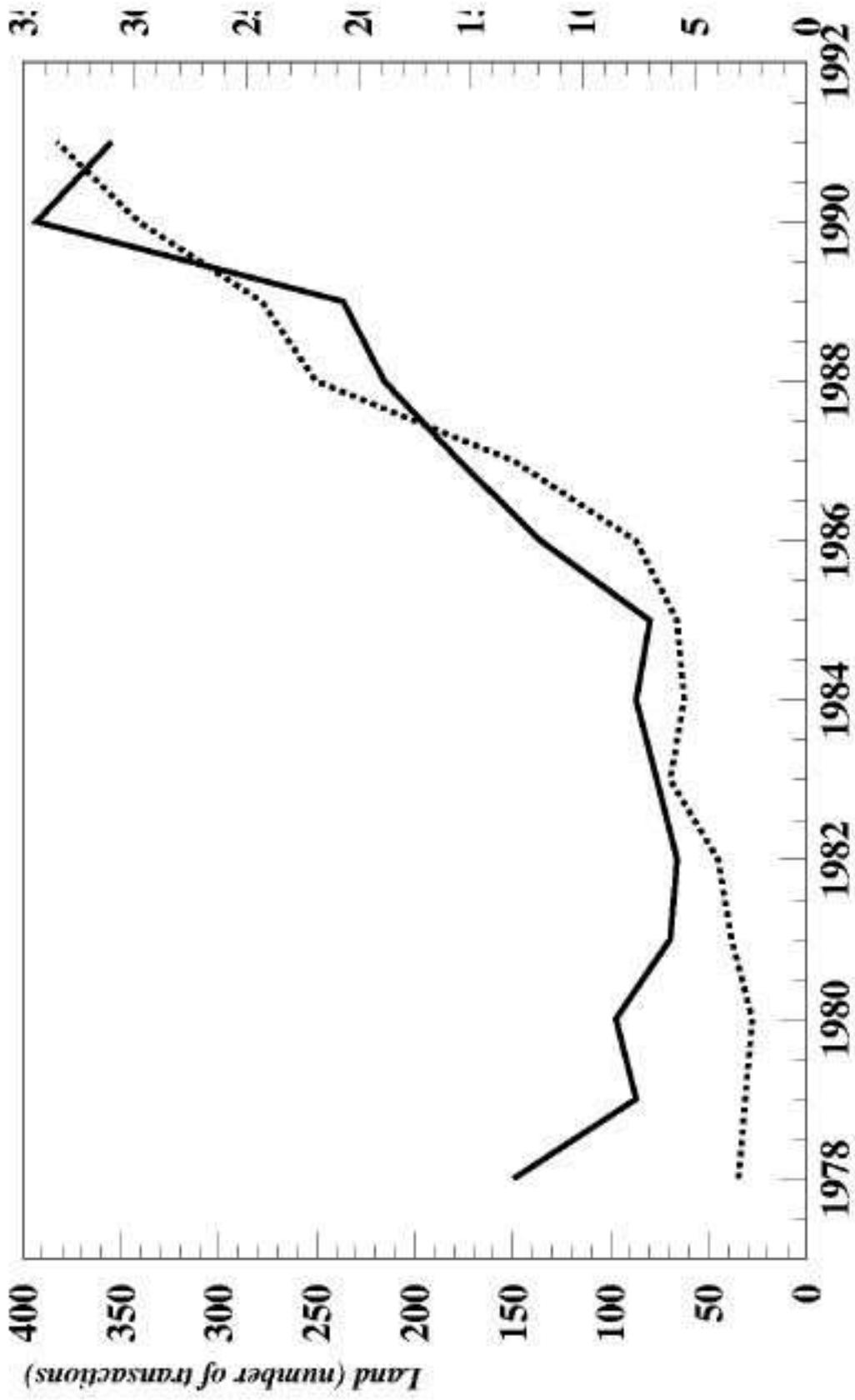
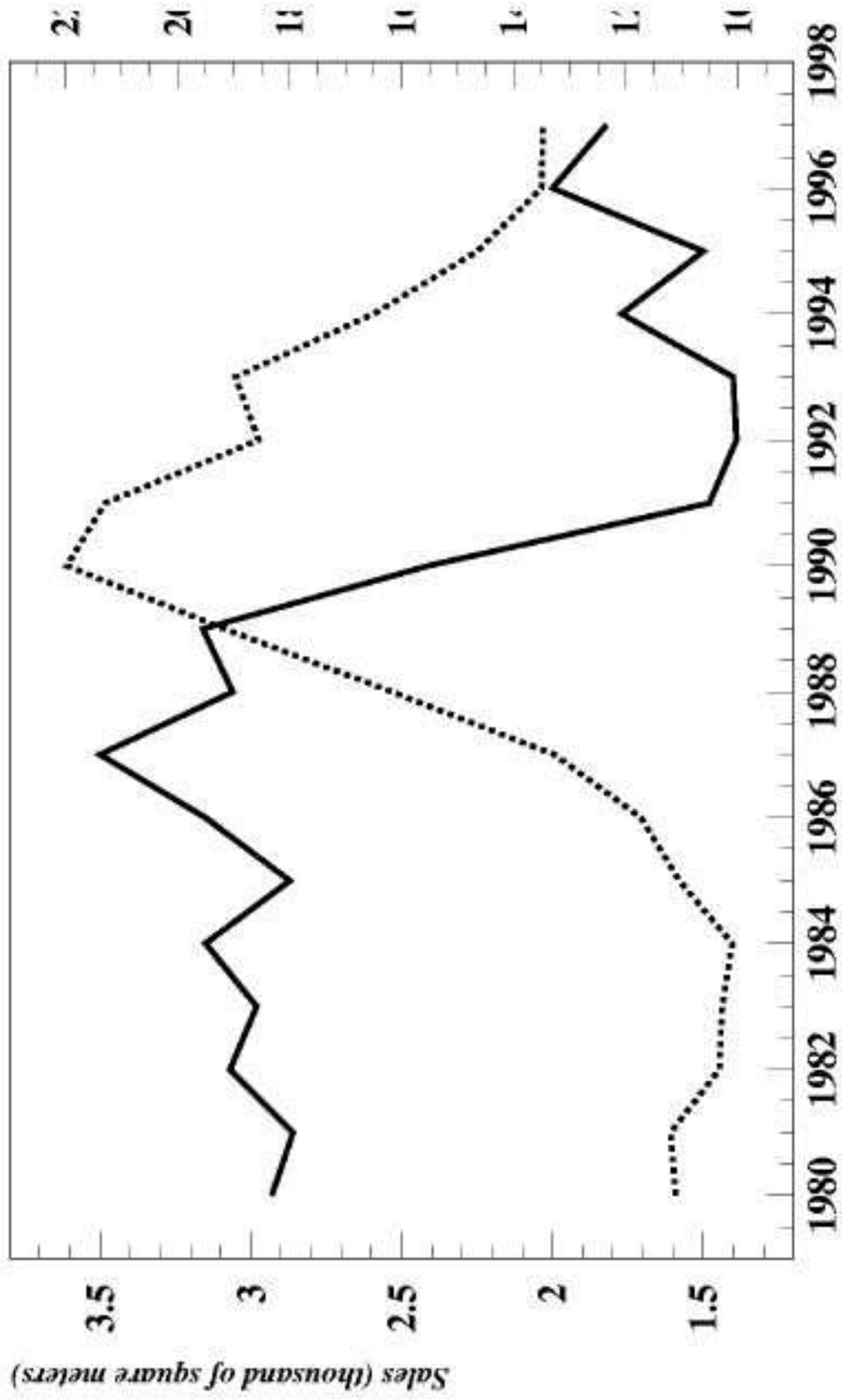
\end{document}